\documentclass[reprint, amsmath, amssymb, aps, prl, superscriptaddress, nolongbibliography, nofootinbib]{revtex4-1}

\usepackage{graphicx}
\usepackage{dcolumn}
\usepackage{xspace}
\usepackage{bm}
\usepackage[normalem]{ulem}
\usepackage{orcidlink}
\usepackage{newtxtext,newtxmath}
\usepackage{amsmath}
\usepackage{amssymb}
\usepackage{tabularx}
\usepackage{bbold}

\newcommand{\nv}{\hat{\bf n}}
\newcommand{\planck}{{\sl Planck}\xspace}
\newcommand{\des}{DES Y3\xspace}
\newcommand{\metall}{Z}
\newcommand{\lMc}{\log_{10}M_c}
\newcommand{\lTAGN}{\log_{10}T_{\rm AGN}}
\newcommand{\omegab}{\Omega_{\text{b}}}

\newcommand{\omegam}{\Omega_{\text{m}}}

\newcommand{\protm}{m_{\text{p}}}

\DeclareMathOperator{\bnd}{bnd}

\newcommand{\fig}[3]{
	\begin{figure}
		\centering
		\includegraphics[width=#2\textwidth]{#1.pdf}
        \vspace{-15pt}
		\caption{#3}
		\label{fig:#1}
        \vspace{-10pt}
\end{figure}}

\newcommand{\widefig}[3]{
	\begin{figure*}
		\centering
		\includegraphics[width=#2\textwidth]{#1.pdf}
        \vspace{-10pt}
		\caption{#3}
		\label{fig:#1}
        \vspace{-10pt}
\end{figure*}}

\newcommand{\mnras}{Mon. Not. R. Astron. Soc.} %
\newcommand{\jcap}{J. Cosmol. Astropart. Phys.} %

\newcommand{\oja}{Open J. Astrophys.} 
\newcommand{\aap}{Astron. Astrophys.} 

\begin{document}

\title{X-Ray–Cosmic-Shear Cross-Correlations: First Detection and Constraints on Baryonic Effects}
\author{Tassia Ferreira\orcidlink{0000-0002-6731-9329}}\email{mailto:tassia.ferreira@physics.ox.ac.uk}
\affiliation{Department of Physics, University of Oxford, Denys Wilkinson Building, Keble Road, Oxford OX1 3RH, United Kingdom}
\author{David Alonso\orcidlink{0000-0002-4598-9719}}
\affiliation{Department of Physics, University of Oxford, Denys Wilkinson Building, Keble Road, Oxford OX1 3RH, United Kingdom}
\author{Carlos Garcia-Garcia\orcidlink{0000-0001-6394-7494}}
\affiliation{Department of Physics, University of Oxford, Denys Wilkinson Building, Keble Road, Oxford OX1 3RH, United Kingdom}
\author{Nora Elisa Chisari\orcidlink{0000-0003-4221-6718}}
\affiliation{Institute for Theoretical Physics, Utrecht University, Princetonplein 5, 3584 CC, Utrecht, The Netherlands}

\date{\today}

\begin{abstract}
  We report the first detection, at very high significance ($23\sigma$), of the cross-correlation between cosmic shear and the diffuse x-ray background, using data from the Dark Energy Survey and the ROSAT satellite. The x-ray cross-correlation signal is sensitive to the distribution of the surrounding gas in dark matter halos. This allows us to use our measurements to place constraints on key physical parameters that determine the impact of baryonic effects in the matter power spectrum. In particular, we determine the mass of halos in which feedback has expelled half of their gas content on average to be $\log_{10}(M_c/M_\odot)=13.643^{+0.081}_{-0.12}$, and the polytropic index of the gas to be $\Gamma = 1.231^{+0.015}_{-0.011}$. This represents a first step in the direct use of x-ray cross-correlations to obtain improved constraints on cosmology and the physics of the intergalactic gas.
\end{abstract}

\maketitle

\section{Introduction}\label{sec:intro}
\vspace{-0.5cm}
  The cosmic baryonic component contains invaluable information to advance our understanding of cosmology and astrophysics from large-scale structures (LSSs). Because of the complexity of the astrophysical processes used to probe the cosmic gas, this information is not fully utilised. Understanding the baryonic component is vital to obtain robust cosmological constraints. If not modelled, baryonic feedback processes may significantly bias cosmological parameters extracted from weak lensing data, for cosmic shear alone, and its combination with galaxy clustering (the ``$3 \times 2$-point'' analysis)~\cite{Semboloni:2011fe, Schneider:2018pfw, Chisari:2019tus}.

  In this Letter, we use a novel probe: the cross-correlation between diffuse x-ray maps and cosmic shear data. Hot gas in overdense regions emits in x-rays through bremsstrahlung and collisionally-ionised line emission. As a result, the x-ray emissivity is proportional to the square of the gas density, and to a function $\Lambda(T_{\text{g}},\metall)$, which depends on the gas temperature and chemical composition~\cite{Peterson:2005hr}. This dependence on density and temperature [$\propto n_{\text{g}}^2\Lambda(T_{\text{g}})$] makes it possible to use x-ray observations to constrain baryonic effects and cosmology. This has been done indirectly using observations of bound gas fractions in x-ray cluster observations~\cite{Jakobs:2017rcd,Schneider:2019xpf,Schneider:2021wds,Grandis:2023qwx}.
  Instead, here we directly explore this cross-correlation to constrain baryonic physics.
  This complements other x-ray cross-correlation studies in the literature~\cite{Hurier:2015ohq,Hurier:2017rgp,Lakey:2019aqt,Hurier:2014upv,Hurier:2015ohq,Shirasaki:2019ndb,Lau:2022dub,Diego:2003cv}.

  The use of this cross-correlation has clear advantages. First, it directly involves the main observable, cosmic shear, where the impact of mismodelled baryonic physics is most relevant. Second, the cross-correlation signal is sourced by the collective emission of all LSSs, making the analysis less sensitive to errors in the modelling of individual objects (only statistical means over the whole halo ensemble are needed). Third, the tomographic nature of cosmic shear data gives insights on the time evolution of the main gas properties. Finally, using the same type of summary statistics (angular power spectra or correlation functions) as in the standard cosmic shear and $3 \times 2$-point analysis allows a straightforward incorporation of the associated measurements. This enables a cosmological analysis that self-consistently constrains cosmology and baryonic effects, propagating all measurement and modelling uncertainties exactly. Here we present a first detection of the cross-correlation between cosmic shear and the soft x-ray background, using data from the ROSAT All-Sky Survey (RASS~\cite{Voges:1999ju}), and the year-3 data release of the Dark Energy Survey (DES Y3~\cite{DES:2020ekd,DES:2020aks}). We show that the cross-correlation signal is detected at very high significance in all redshift bins, providing strong bounds on baryonic effects.

\vspace{-0.4cm}
\section{Data analysis}\label{sec:data}
\vspace{-0.3cm}

  {\bf The Dark Energy Survey.}\enskip We use the cosmic shear data made available with the \des~\cite{DES:2020ekd}. We process the source catalogue following the steps in the DES cosmic shear analysis~\cite{DES:2021vln,DES:2022qpf}. The catalogue is divided into four redshift bins, and galaxy ellipticities are corrected for additive and multiplicative shape measurement biases calculated from the weighted average ellipticity and shear response tensor in each bin. After obtaining the source weights and corrected ellipticities, we produce shear maps following the steps in~\cite{Nicola:2020lhi}. We use the HEALPix pixelization scheme~\cite{Gorski:2004by} with resolution parameter $N_{\rm side}=1024$, corresponding to pixels of size $\delta\theta\simeq3.4'$. When interpreting the power spectrum measurements described below, we use the redshift distributions employed in the \des analysis~\cite{DES:2020ebm}.

  {\bf ROSAT.} \enskip We use data from the RASS~\cite{Voges:1993,Voges:1999ju}, covering the full celestial sphere. ROSAT is currently the only publicly available all-sky x-ray survey, before the data taken by eROSITA~\cite{eROSITA:2012lfj} are released. We produce maps of the photon count rate density (i.e. number of photons observed per unit time and solid angle) in the observed energy range $E\in[0.5,2]$ keV. For this, we use the photon table available at the German Astrophysics Virtual Observatory~\cite{gavo}, and of the merged exposure maps from the RASS data release \footnote{\url{https://heasarc.gsfc.nasa.gov/FTP/rosat/data/pspc/processed_data/rass/release/}.}. Exposure maps were reprojected onto a HEALPix map with resolution $N_{\rm side}=1024$, and the photon count rate density map was constructed by binning all photons in the energy range onto these pixels and dividing by the corresponding exposure and pixel area. We verified that the photon count and exposure maps thus constructed matched those made available by the Centre d’Analyse de Donn\'ees Etendue \footnote{\url{http://cade.irap.omp.eu/dokuwiki/doku.php?id=rass}.}.

  {\bf Power spectra and covariances.} \enskip We estimate the cross-\\ correlation between the shear and x-ray maps using the pseudo-$C_\ell$ estimator~\cite{Hivon:2001jp,Alonso:2018jzx} as implemented in {\tt NaMaster}~\cite{Alonso:2018jzx}. The treatment of the cosmic shear data closely followed the description in~\cite{Nicola:2020lhi}. To account for the inhomogeneous noise properties in ROSAT, the mask used for the photon count rate map in the pseudo-$C_\ell$ estimator was proportional to the exposure map. We masked pixels with exposures below $100$s, and imposed a Galactic mask constructed by thresholding the dust extinction map of~\cite{Schlegel:1997yv} to minimise potential contamination from Galactic foregrounds (see~\cite{Koukoufilippas:2019ilu}). The covariance matrix of our measurements was constructed analytically with the methods introduced in~\cite{Garcia-Garcia:2019bku,Nicola:2020lhi}, using the power spectra directly computed from the data. We verified that our code was able to reproduce the cosmic shear angular power spectra and covariance matrix published by DES~\cite{DES:2022qpf}. The power spectra were calculated in the same $\ell$ bins used in~\cite{Garcia-Garcia:2021unp}, only considering angular scales in the range $30<\ell<2N_{\rm side}$, resulting in  24 bins per $C_\ell$. We incorporated the effects of marginalising over uncertainties in the DES redshift distributions and on the multiplicative bias into the power spectrum covariance using the method described in~\cite{Ruiz-Zapatero:2023tnt}. Although our analytical covariance matrix accounts only for Gaussian contributions, we verified that including non-Gaussian terms (one-halo trispectrum and supersample covariance) changes$\chi^2$ values by less than 2\%.

\vspace{-0.4cm}
\section{Theoretical model}\label{sec:model}
\vspace{-0.3cm}

  {\bf X-ray count rate density maps.} \enskip The observed count rate density of x-ray photons along the directional unit vector $\nv$ is given by
  \begin{equation}
    {\rm CR}(\nv)=\int\frac{d\chi}{4\pi(1+z)^4}\,n_{\text{e}}(\chi\nv)\,n_{\text{H}}(\chi\nv)J(T_{\text{g}},\metall)\,,
  \end{equation}
  where $\chi$ is the comoving radial distance, $n_{\text{e}}$ and $n_{\text{H}}$ are the physical number densities of free electrons and hydrogen nuclei, and $T_{\text{g}}$ and $\metall$ are the temperature and metallicity of the plasma. $J(T_g,\metall)$ is the x-ray emissivity of the gas convolved with the instrument response function and integrated over the observed energy range. I.e.:
  \begin{align}\nonumber
    J(T_{\text{g}},\metall)\equiv\int_{E_{\rm min}}^{E_{\rm max}}& dE_o\int_0^\infty dE\, \mathcal{M}(E_o|E)A(E)\\
    &\frac{d\epsilon}{dE}\Big(T_{\text{g}},\metall,E(1+z)\Big)\,.
  \end{align}
  Here, $E$ is the true photon energy in the observer's rest frame, $E_o$ is the energy measured by ROSAT. $\mathcal{M}(E_o|E)$ is the energy redistribution matrix, and $A(E)$ is the effective detector area for photons with energy $E$; these are given by the response matrix files and the ancillary response files, respectively. $d\epsilon/dE$ is the x-ray emissivity (number of photons emitted per unit volume, time, energy, and squared number density), evaluated at the rest-frame energy of the plasma, $E\,(1+z)$. We compute this quantity using the Astrophysics Plasma Emission Code ({\tt APEC}~\cite{Smith:2001he}) as implemented in {\tt PYATOMDB}~\cite{Foster:2012hy}.

  Assuming a fully ionised plasma, $n_{\text{e}}$ and $n_{\text{H}}$ are proportional to the gas mass density $\rho_{\text{g}}$: $n_{\text{H}}=\rho_{\text{g}}X_{\text{H}}/\protm$, and $n_{\text{e}}=\rho_{\text{g}}(1+X_{\text{H}})/(2\protm)$, where $X_{\text{H}}=0.76$ is the hydrogen mass fraction, fixed at the primordial abundance. The x-ray map is thus sensitive to the physical properties of the gas (density, temperature, metallicity). We assume a metallicity $\metall=0.3\metall_\odot$, where $\metall_\odot$ is the solar metallicity~\cite{Mernier:2018tzs}.

  {\bf A hydrodynamical halo model.} \enskip To describe the distribution of mass and gas within halos we use a halo model, strongly inspired by~\cite{Mead:2020qgo} (M20 hereon). Following M20, we consider contributions to the total halo mass from cold dark matter (CDM), gravitationally bound gas, ejected gas, and stars.

  As in M20, we model the bound gas assuming hydrostatic equilibrium and $P\propto\rho^\Gamma$. This results in the so-called ``KS'' density profile~\cite{Komatsu:2001dn}, with $\Gamma$ determining its scale dependence. The profile normalisation is fixed by requiring the total bound gas mass to follow the relation~\cite{Schneider:2015wta} $f_{\bnd}(M) = (\omegab/\omegam)/[(1 + (M_c/M)^{\beta}]$.
  The mass scale $M_c$ (which we refer to as the ``halfway mass'') is the halo mass at which half of the gas mass has been ejected. We fix $\beta = 0.6$~\cite{Schneider:2015wta,Mead:2020qgo}. The stellar component is concentrated at the halo centre and modelled as in Eq. (27) of M20 and~\cite{Fedeli:2014hca}. The rest of the gas mass (adding up to the cosmic mean) is assumed to be in the form of ejected gas. We model the scale dependence of this component following~\cite{Schneider:2015wta}, using a Gaussian profile with a characteristic ejection scale controlled by a free parameter $\eta_b$ [see Eq. (2.25) of~\cite{Schneider:2015wta}]. The CDM component is assumed to follow a Navarro-Frenk-White profile~\cite{NFW} (NFW). Finally, we account for the effects of clumping using the model of~\cite{Battaglia:2014cga}. While our data are not sensitive to the details of the ejected gas distribution, as it is sufficiently diffuse and cold to avoid detection via x-rays, the ``clumpiness'' of the bound gas can have a significant effect on the signal, and its model should be carefully calibrated.

  The temperature of the bound and ejected gas is modelled as in M20. The bound gas temperature follows the scale dependence predicted by the polytropic equation of state and the NFW gravitational potential. We assume the gas to be close to virialised, with a free parameter $\alpha_T$ describing departures from a perfect virial relation [for which $\alpha_T=1$; see Eq. (39) of M20]. As in M20 we fix the temperature of the ejected gas to that of the warm-hot IGM $T_w=10^{6.5}\,{\rm K}$.

  The model thus depends on 4 free parameters, which we vary assuming flat priors: $\lMc\in[13,15]$, $\Gamma\in[1.1, 1.5]$, $\alpha_T\in[0.5, 2.0]$, and $\eta_b\in[0.2, 2.0]$. To transform this model into predictions for angular power spectra, we use the general formalism described in e.g.~\cite{Koukoufilippas:2019ilu}. We use a halo mass definition with overdensity parameter $\Delta=200$ times the critical density, the mass function and halo bias prescriptions of~\cite{Tinker:2010my}, and the concentration-mass relation of~\cite{Duffy:2008pz}. We note that we have ignored the impact of intrinsic alignments (IAs), since the \des analysis did not find significant evidence for this contribution~\cite{DES:2021vln}. We verified that including the IAs with an amplitude $A_{\rm IA}=1$ changed the results by less than 1 standard deviation, as consistent with KiDS~\cite{KiDS:2020suj}. A more detailed description of the model, and the role of its parameters, will be provided in future work. All theoretical calculations were carried out using the Core Cosmology Library~\cite{LSSTDarkEnergyScience:2018yem}. We keep all cosmological parameters fixed to the best-fit $\Lambda$CDM values found by \planck~\cite{Planck:2018vyg}, since it is not possible to simultaneously constrain the cosmological and baryonic physics parameters using only the x-ray cross-correlation studied here \footnote{For instance, both the halfway mass $M_c$ and $\sigma_8$ mainly affect the overall amplitude of the cross-spectrum.}. Together with IA parameters, the cosmology should be marginalised over in a joint analysis with the cosmic shear autocorrelations.

\vspace{-0.4cm}
\section{Results}\label{sec:res}
\vspace{-0.3cm}
  \fig{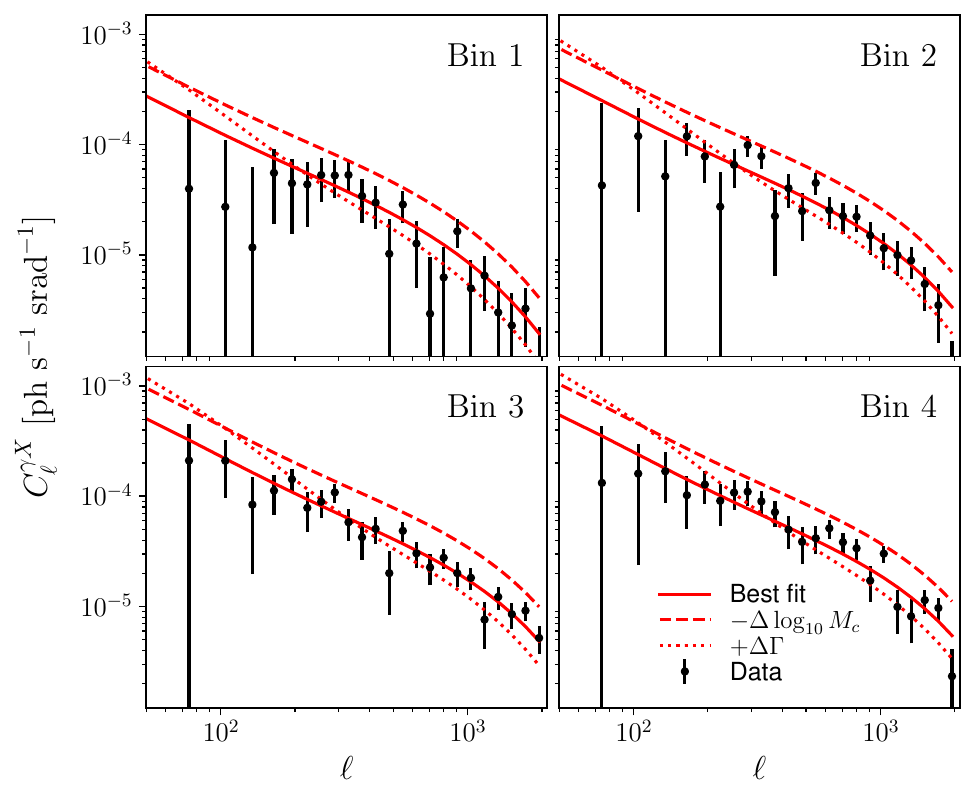}{0.49}{
  Cross-correlations between the ROSAT count rate density map and the \des cosmic shear data in four redshift bins for the scales used in the analysis ($30<\ell<2N_{\rm side}$). The red solid line shows our best-fit prediction, with the dashed and dotted lines showing predictions deviating from this model by $5\sigma$ in $\lMc$ and $\Gamma$, respectively.}

  {\renewcommand{\arraystretch}{1.5}
  \setlength{\tabcolsep}{5pt}
  \begin{table*}
  \begin{center}
    \begin{tabular}{|c|ccccc|ccc|cc|}
      \hline
      Data & $\log_{10}M_c$ & $\Gamma$ & $\chi^2_{\rm min}$ & PTE & SNR & $\lTAGN$ & $\chi^2_{\rm min}$ & PTE & $\chi^2_B$ & PTE$_B$\\
      \hline \hline
      \, Bin 1 \, & $14.36^{+0.19}_{-0.37}$ & $1.203^{+0.042}_{-0.034}$ & 16.5 & 0.68 & 6.0 & $8.029 \pm 0.028$ & 17.0 & 0.81 & 13.5 & 0.97\\
      \, Bin 2 \, & $14.179^{+0.097}_{-0.26}$ & $1.214^{+0.036}_{-0.020}$ & 24.2 & 0.23 & 11.1 & $8.005 \pm 0.016$ & 25.9 & 0.30 & 17.3 & 0.83\\
      \, Bin 3 \, & $14.184^{+0.091}_{-0.26}$ & $1.206^{+0.034}_{-0.019}$ & 19.0 & 0.52 & 15.6 & $7.998 \pm 0.012$ & 19.1 & 0.69 & 21.0 & 0.64\\
      \, Bin 4 \, & $14.090^{+0.068}_{-0.20}$ & $1.216^{+0.030}_{-0.016}$ & 23.8 & 0.25 & 15.1 & $7.987 \pm 0.012$ & 25.9 & 0.31 & 26.9 & 0.31\\
      \, All bins \, & $14.059^{+0.065}_{-0.13}$ & $1.224^{+0.018}_{-0.011}$ & 80.1 & 0.81 & 23.4 & $7.998 \pm 0.008$ & 82.1 & 0.82 & 78.7 & 0.90\\
      \hline
    \end{tabular}
  \end{center}
  \vspace{-10pt}
  \caption{Parameter bounds, best-fit $\chi^2$, associated PTE, and detection significance from each redshift bin individually (first four rows) and combined (last row). The first five columns show results for the two parameters constrained by the data $\{\lMc,\Gamma\}$ in the four-parameter model (varying $\{\lMc,\Gamma,\alpha_T,\eta_b\}$). The next three columns show results for the single-parameter model of M20 using the AGN feedback parameter $\lTAGN$. The last two columns show $\chi^2$ and PTE for the corresponding $B$-mode power spectra.}\label{tab:results}
  \end{table*}}

  \widefig{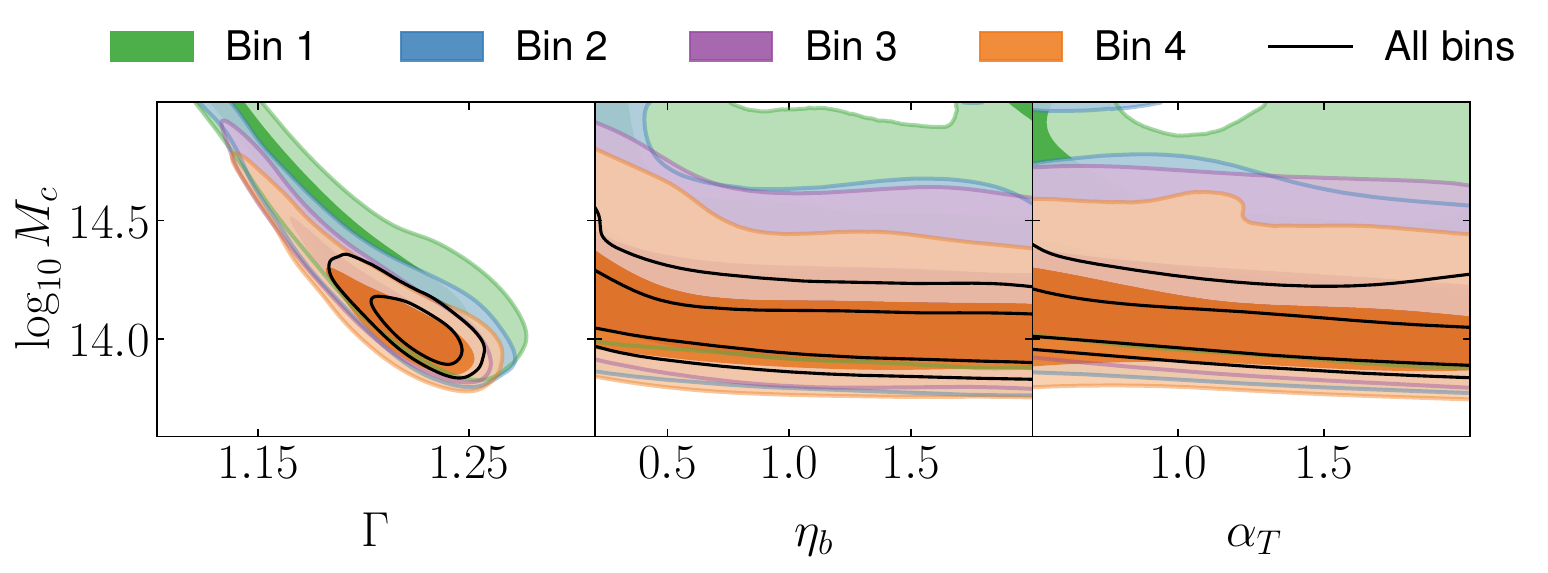}{0.9}{Constraints on the four parameters of the hydrodynamical halo model used in this Letter. Results are shown for each of the four \des redshift bins independently (shaded coloured contours) and for the combination of all bins (solid black line). Our data are most sensitive to the density of bound gas and constrains the halo mass at which half of the gas content has been ejected ($\lMc$), which controls the overall amplitude of the density profile, and the polytropic index ($\Gamma$), which controls its scale dependence. The departure from the virial temperature relation $\alpha_T$ and the scale of the ejected gas ($\eta_b$) are largely unconstrained.}

  {\bf Measurement and detection significance.} We estimate the power spectra between the ROSAT count rate density map, and cosmic shear $E$-mode in the four DES redshift bins. The measurements are shown in Fig. \ref{fig:allcl2x2}, together with the best-fit model prediction. The detection significance is calculated as SNR$=\sqrt{\chi_{\rm min}^2-\chi^2_0}$, where $\chi_{\rm min}^2$ and $\chi^2_0$ are the $\chi^2$ for the best-fit model and for a null model respectively. The goodness of fit of the model is quantified in terms of the probability to exceed (PTE) for the best-fit $\chi^2$ assuming a $\chi^2$ distribution with $N_{\rm d.o.f.}=N_d-N_p$ degrees of freedom, where $N_d$ is the number of data points and $N_p=4$ is the number of free parameters of the model. Table \ref{tab:results} lists the best-fit $\chi^2$s, PTEs, and SNRs for each redshift bin and for the combination of all measurements. The signal is detected at high significance ($S/N>7$) in all redshift bins, with a total signal-to-noise ratio of SNR$=23.4$. The model provides a good fit to the data in all bins individually and jointly, with PTE above 0.3. The joint best-fit model is also a good fit to each power spectrum, with PTE ranging between 0.29 and 0.72.

  To validate these measurements, we perform a null test, evaluating the detection significance of the correlation between ROSAT and the cosmic shear $B$-modes. As expected, we find these measurements to be compatible with zero in all bins (the associated $\chi^2$ values and probabilities to exceed are listed in Table \ref{tab:results}).
  
  {\bf Constraints on baryonic effects.} To constrain the four parameters of our model, $\{\lMc,\Gamma,\alpha_T,\eta_b\}$, we assume a Gaussian likelihood, which we sample with the Metropolis-Hastings Monte Carlo Markov Chain (MCMC) algorithm in {\tt COBAYA}~\cite{Torrado:2020dgo}.
  We present parameter constraints by quoting the peak of the marginalised 1D distribution, together with its $68\%$ asymmetric confidence-level interval. Fig. \ref{fig:rectangle_main} shows 2D contours for the four parameters for each redshift bin (coloured shaded contours), and for the total data vector (black contours).
  
  As anticipated, the x-ray data are most sensitive to the properties of the bound gas and constrained both the halfway mass scale $\lMc$ (which controls the amplitude of the cross-correlation) and the polytropic index $\Gamma$ (which controls its scale dependence). We are mostly insensitive to variations concerning the virial temperature (encoded in $\alpha_T$) 
  and the scale dependence of the ejected gas profile (encoded in $\eta_b$), since its low temperature renders it invisible to x-rays. With the full data vector, our constraints on $\lMc$ and $\Gamma$ are
  \begin{equation}
    \lMc=14.059^{+0.065}_{-0.13},\hspace{12pt} \Gamma=1.224^{+0.018}_{-0.011}.
  \end{equation}
  These results are in good agreement with those found in cosmic shear analyses when including the impact from baryonic effects~\cite{DES:2022eua,Amon:2022azi,Kilo-DegreeSurvey:2023gfr,Preston:2023uup,Arico:2023ocu} and when considering x-ray measurements of gas fractions~\cite{Schneider:2019xpf,Grandis:2023qwx}. Our constraints are nevertheless significantly tighter, particularly in the case of $\lMc$, which is the main parameter regulating the impact of baryonic effects in weak lensing data~\cite{Arico:2023ocu}. Table \ref{tab:results} shows the constraints found for each redshift bin. We find reasonable agreement between bins, with a mild trend for $\lMc$ to decrease with redshift, and no clear trend in the case of $\Gamma$.

  {\bf Constraints on single-parameter models.} 
  Since there have been attempts at modelling baryonic effects on cosmic shear analysis with a single free parameter, it is thus interesting to explore whether our measurements can be described by these simpler models. In particular, we loosely follow the prescription of M20, which was applied to the joint analysis of cosmic shear and thermal Sunyaev-Zel'dovich (tSZ) data in~\cite{Troster:2021gsz}.

  In~\cite{Troster:2021gsz}, the authors fitted the density and pressure power spectra in the BAHAMAS suite of hydrodynamical simulations~\cite{McCarthy:2016mry} to the halo model described above and connected the various parameters of the model to a single ``AGN temperature'' parameter $\lTAGN$, which regulates the strength of feedback in the simulations. Using the results shown in Table 2 of M20, we reparametrise our model as a function of $\lTAGN$ and use our data to constrain this parameter. It is worth noting that our model assumes a Gaussian profile for the ejected gas, whereas~\cite{Troster:2021gsz} set the one-halo contribution from the ejected gas to zero. Since our data are not sensitive to the ejected gas parameters, we fix $\eta_b=0.5$ following~\cite{Schneider:2015wta}.

  The constraints on $\lTAGN$ are listed in Table \ref{tab:results}. We find $\lTAGN=7.998\pm0.008$ from our combined data vector, which is in excellent agreement with both the BAHAMAS model~\cite{Troster:2021gsz} and the constraints found by~\cite{DES:2022eua,Arico:2023ocu} from the joint analysis of cosmic shear and tSZ data using this model. The joint constraint is also in reasonable agreement with those found for individual redshift bins, and the model provides a good fit to the data in all cases. Although there seems to be some qualitative evidence of evolution in $\lTAGN$, this is not easy to quantify, given the large overlap between the lensing kernels of all bins. The trend might also be purely an artifact of the $\lTAGN$ parametrisation or depend on the impact from intrinsic alignments at low redshifts, which were neglected.

\vspace{-0.4cm}
\section{Discussion}\label{sec:disc}
\vspace{-0.3cm}

  We conducted the first analysis of the angular power spectrum between x-ray count maps and cosmic shear data. The signal is detected at very high significance, allowing us to place strong constraints on the free parameters of a hydrodynamical halo model describing the gas content of halos. In particular, we constrain the halfway mass $\lMc$ and the polytropic index $\Gamma$, which govern the abundance of bound and ejected gas, and the scale dependence of the former. Both parameters, particularly $\lMc$, play significant roles in modelling and mitigating the impact of baryonic effects in the matter power spectrum. This is a major source of uncertainty in cosmic shear analyses, preventing the use of smaller scales to improve cosmological constraints.

  \fig{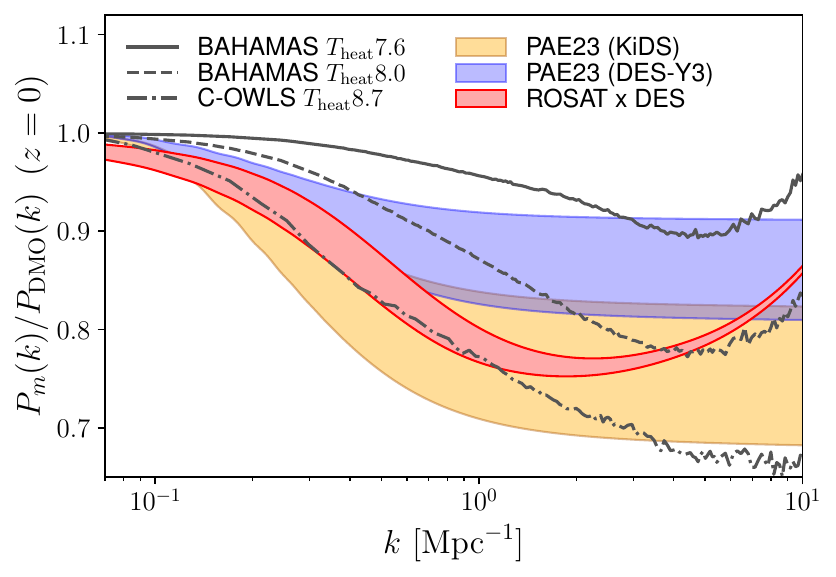}{0.49}{
  Suppression of the matter power spectrum due to baryonic effects. The solid, dashed and dot-dashed lines show results for various hydrodynamical simulations~\cite{Brun:2013yva,McCarthy:2016mry}. The blue and yellow bands show the level of suppression needed to reconcile KiDS 1000 and DES Y3 cosmic shear data with \planck~\cite{Preston:2023uup}. The red band is the 1$\sigma$ bounds obtained from the ROSAT-DES cross-correlations within our fiducial four-parameter model, converging to unity at larger scales.}

  As shown, x-rays are an excellent probe for this parameter as it traces the hot gas inside halos and thus provides strong bounds for its properties. For reference, Fig. \ref{fig:baryon_boost} shows the level of suppression of the matter power spectrum predicted by various hydrodynamical simulations~\cite{Brun:2013yva,McCarthy:2016mry}, the level needed to reconcile cosmic shear data with CMB data from \planck (as quantified in~\cite{Preston:2023uup}), and the bounds on this suppression imposed by our constraints on $\lMc$ and $\Gamma$. It is worth emphasizing that these bounds are only indirect evidence for this suppression, based on the constraints on the distribution of gas arising from the shear--x-ray cross-correlation. Our constraints straddle the levels of suppression preferred by DES and KiDS, which may have an impact in solving the so-called ``$S_8$ tension''~\cite{Amon:2022azi,Preston:2023uup, Arico:2023ocu}. The x-ray--cosmic-shear cross-correlation presented here can thus be used to calibrate the impact of baryonic effects. In contrast with previous approaches to using x-ray measurements of cluster samples for similar purposes~\cite{Jakobs:2017rcd,Schneider:2019xpf,Schneider:2021wds,Grandis:2023qwx}, the cross-correlation approach allows for a joint analysis with any other cosmic shear correlations (e.g. $3 \times 2$-point analyses), consistently and naturally propagating all uncertainties and accounting for the covariance between probes. Residual degeneracy between cosmological and hydrodynamical parameters could be broken by including cross-correlation with Sunyaev-Zel'dovich Compton-$y$ maps, given their complementary sensitivity to gas density and temperature. We leave such analysis for future work. Alternatively, $X$-ray cross-correlations could be used in the calibration of baryonic physics in hydrodynamical simulations~\cite{McCarthy:2016mry}.

  Given the high sensitivity of these measurements, a careful validation of the theoretical model used to describe them, is crucial before the methodology presented here can be used in cosmological analyses. Here we utilised a simple halo model parametrisation, which is likely subject to several sources of uncertainty. These include the mismodelling of the one- to two-halo transition regime, the idealised nature of the various halo profiles used, the hydrostatic mass bias (used to rescale  halo masses to account for non-thermal contributions to the pressure)~\cite{Schneider:2018pfw}, and the stochastic relationship between halo mass and gas properties that could affect the one-halo contribution. Other x-ray-specific systematics, such as the detailed chemical composition of the IGM or the effects of clumping, will also require a more careful treatment. Finally, we kept all cosmological parameters fixed to their best-fit \planck values and neglected the contribution from intrinsic alignments. As argued here, both assumptions would not affect the qualitative conclusions made, regarding the potential of cross-correlations with the diffuse x-ray background to constrain baryonic physics. However, they would need to be addressed in a joint cosmological analysis involving e.g. cosmic shear autocorrelations.

  This Letter paves the way for the use of cross-correlations between cosmic shear and maps of the diffuse x-ray background to address central questions in the path to obtain robust cosmological constraints from large-scale structure data. The inclusion of this probe in the joint analysis of future weak lensing surveys (e.g. the Vera Rubin Observatory~\cite{LSSTScience:2009jmu} and Euclid~\cite{Refregier:2010ss}) with ongoing x-ray missions, such as eROSITA~\cite{eROSITA:2012lfj}, will be instrumental to improve the precision of the associated cosmological constraints and their robustness to astrophysical uncertainties.

\begin{acknowledgments}
  \noindent
  {\bf Acknowledgements.}
  We thank Raul Angulo, Cyrille Doux, Deborah Paradis, Aurora Simionescu, Tilman Tr\"oster, Matteo Zennaro and Giovanni Aric\'o for useful comments and discussions. We also thank the anonymous journal referees, whose comments helped improve the quality of this Letter. T.F. is supported by a Royal Society Newton International Fellowship. D.A. and C.G.G. acknowledge support from the Beecroft Trust. This research was funded in whole, or in part, by the Dutch Research Council (NWO) 24.001.027. We made extensive use of computational resources at the University of Oxford Department of Physics, funded by the John Fell Oxford University Press Research Fund. For the purpose of open access, a CC BY public copyright license is applied to any Author Accepted Manuscript version arising from this submission.
\end{acknowledgments}

\bibliography{references}

\end{document}